\documentclass[12pt]{article}

\pdfoutput=1

\usepackage{graphics}
\usepackage{epsfig}

\usepackage{amsfonts}
\usepackage{amssymb}
\usepackage{amsmath}
\usepackage{dsfont}
\usepackage{multirow}
\usepackage{latexsym}
\usepackage{verbatim}
\usepackage{mcite}
\usepackage{cite}
\usepackage{units}
\usepackage[all]{xy}
\usepackage[usenames,dvipsnames,table]{xcolor}
\usepackage{cancel}
\usepackage{slashed}
\usepackage{hyperref}
\usepackage{physics}
\usepackage{breqn}

\def\beq{\begin{equation}}
\def\eeq{\end{equation}}
\def\beqa{\begin{eqnarray}}
\def\eeqa{\end{eqnarray}}
\newcommand{\f}{\frac}

\def\bfone{\relax{\rm 1\kern-.35em 1}}

\def\vx{{\mathbf{x}}}

\def\vp{{\mathbf{p}}}
\def\vq{{\mathbf{q}}}
\def\vk{{\mathbf{k}}}

\newcommand{\cD}{{\cal D}}

\newcommand{\cO}{{\cal O}}

\newcommand{\cU}{{\cal U}}

\newcommand{\be}{\begin{equation}}
\newcommand{\ee}{\end{equation}}
\newcommand{\ben}{\begin{displaymath}}
\newcommand{\een}{\end{displaymath}}
\newcommand{\bea}{\begin{eqnarray}}
\newcommand{\eea}{\end{eqnarray}}

\newcommand{\bean}{\begin{eqnarray*}}
\newcommand{\eean}{\end{eqnarray*}}

\DeclareMathAlphabet{\mathpzc}{OT1}{pzc}{m}{it}

\begin{document}
\pagestyle{plain}

\begin{titlepage}
\begin{flushright}
\end{flushright}

\bigskip

    \begin{center}
		{\Large \bf{On the Entanglement Entropy in Gaussian cMERA
	}}
	
	\vskip 1.5cm

	{\bf J. J. Fern\'andez-Melgarejo$^\spadesuit$\footnote{melgarejo@at@um.es}\, and J. Molina-Vilaplana$^\clubsuit$\footnote{javi.molina@at@upct.es}}
	
\begin{center}
${}^\spadesuit${\it Departamento de F\'isica, Universidad de Murcia,}\\
{\it Campus de Espinardo, 30100 Murcia, Spain}

\vspace*{0.5cm}

${}^\clubsuit${\it Universidad Polit\'ecnica de Cartagena,}\\
{\it Calle Dr. Fleming, S/N, 30202 Cartagena, Murcia}
\end{center}


\vspace{1cm}

	\today
	
\vspace{1cm}

    \end{center}

\begin{abstract}
The continuous Multi Scale Entanglement Renormalization Anstaz (cMERA) consists of a variational method which carries out a real space renormalization scheme on the wavefunctionals of quantum field theories.
In this work we calculate the entanglement entropy of the half space for a free scalar theory through a Gaussian cMERA circuit. We obtain the correct entropy written in terms of the optimized cMERA variational parameter, the local density of disentanglers.
Accordingly, using the entanglement entropy production per unit scale, we study local areas in the bulk of the tensor network in terms of the differential entanglement generated along the cMERA flow. This result spurs us to establish an explicit relation between the cMERA variational parameter and the radial component of a dual AdS geometry through the Ryu-Takayanagi formula.
Finally, based on recent formulations of non-Gaussian cMERA circuits, we argue that the entanglement entropy for the half space can be written as an integral along the renormalization scale whose measure is given by the Fisher information metric of the cMERA circuit. Consequently, a straightforward relation between AdS geometry and the Fisher information metric is also established.

\end{abstract}

\end{titlepage}

\tableofcontents

\section{Introduction}
Entanglement is a key feature to characterize quantum systems. The best known measure of it, entanglement entropy, has been used in a wide range of fields such as condensed matter physics, high energy theory and gravitational physics (see \cite{Nishioka:2018khk} and references therein). Given a system described by a quantum state, for an observer having access only to a subregion $A$ of the total system, all physical predictions are given in terms of the reduced density matrix $\rho_A$. The entanglement entropy  measures the amount of missing information about the total system for this observer, and is given by the von Neumann entropy of the reduced density matrix $\rho_A$, \emph{e.g.},
\begin{equation}
S_A= - {\rm Tr}_A\, \rho_{A} \log \rho_{A}\, .
\end{equation}

In quantum field theory (QFT), computing $S_A$ has shown to be an extraordinarily difficult task. Noteworthily, in the context of the AdS/CFT \cite{Maldacena:1997re, Gubser:1998bc, Witten:1998qj} the entanglement entropy can be computed using one of the central entries in the holographic dictionary, the Ryu–Takayanagi formula, \cite{Ryu:2006bv, Faulkner:2013ana},
\begin{align}
S_A=\frac{{\rm Area}(\gamma_A)}{4\, G_N^{(d+2)}}
\, ,
\label{eq:rt_formula}
\end{align}
which quantifies the entanglement entropy $S_A$ of a region $A$ in a $(d+1)$-QFT admitting a $(d+2)$-gravity dual. Here, $\gamma_A$ is a codimension-2 static minimal surface in AdS$_{(d+2)}$ anchored to the boundary of the region $A$.

The holographic formula for the entanglement entropy \eqref{eq:rt_formula} allows to compute the entanglement entropy in a QFT from the dual bulk geometry. Being the AdS/CFT  a duality between theories, it seems reasonable to think that analyzing the entanglement structure of concrete states in a QFT, one would be able to infer the dual bulk geometries related to these states. Interestingly, this strategy has been graciously revealed in terms of tensor networks, concretely in terms of the Multi-scale Entanglement Renormalization Ansatz (MERA) \cite{Swingle:2009bg}. A MERA tensor network \cite{Vidal:2007hda}, implements a real space renormalization group on the wavefunction of a quantum many body system. A continuous version of MERA (cMERA) has been proposed for free field theories \cite{Haegeman:2011uy, Nozaki:2012zj} and more recently for interacting field theories \cite{Cotler:2018ehb, Cotler:2018ufx, Fernandez-Melgarejo:2019sjo, Fernandez-Melgarejo:2020fzw}. In \cite{Nozaki:2012zj}, to make the connection of cMERA with the AdS/CFT more precise, authors proposed to think in terms of the Fisher information metric defined via quantum distances. Namely, they define the holographic radial component of a dual metric by considering the overlap between states that infinitesimally differ in the renormalization scale of cMERA. However, a more refined  proposal would require to define local areas in the bulk of the tensor network in terms of the differential entanglement generated along the cMERA flow, and see if these local areas can be mapped into minimal areas in AdS spacetimes.
 
Continuing this line of thought, in this paper we have obtained the entanglement entropy for the half-space of a free scalar theory in a Gaussian cMERA tensor network as a function of the  local density of disentanglers, the variational parameter defining the tensor network. This result, which was conjectured in \cite{Nozaki:2012zj} based on an estimation of the entropy in the discrete version of MERA, spurs us to broaden our analysis when other (non-)Gaussian cMERA circuits with additional disentanglers are considered. In particular, we observe that the entanglement entropy must be computed in cMERA through the Fisher information metric $g_{uu}$, which is the avatar of the bond dimension in the discrete version of the MERA circuit.

In addition, our result explicitly shows how the infinitesimal change in the entropy can be cast in terms  of the differential contribution to the area of a minimal surface in a dual AdS geometry. As a result, the dual geometry is defined in terms of the variational parameters of the tensor network through a computation of the entanglement entropy in a QFT state. 

The paper is structured as follows. In Section \ref{sec:ee_qft} we review the obtaining of the entanglement entropy of half space in quantum field theory. Then, after briefly introducing the Gaussian cMERA formalism in Section \ref{sec:cMERA}, in Section \ref{sec:ee_cmera} we study the entanglement entropy in cMERA. In particular, we provide an expression for the entropy as a function of the variational parameter. In Section \ref{sec:holo} we elaborate on the relation of this expression with the Ryu-Takayanagi formula and establish an explicit relation between the AdS metric and the cMERA variational parameter. Finally, we discuss our results and explain our conclusions in Section \ref{sec:discussion}.

\section{Entanglement Entropy of half-space in QFT}\label{sec:ee_qft}
A standard method for the calculation of the entanglement
entropy in a field theory is the {\it replica trick}. 
To illustrate this, and following \cite{Solodukhin:2011gn}, let us consider a quantum field $\psi(X)$ in a $(d+1)$-dimensional spacetime and choose the Cartesian coordinates $X^\mu=\lbrace\tau,x, x_{\perp}^i\rbrace$ with $i=1,..,d{-}1$, where $\tau$ is Euclidean time, such that a surface $A_\perp$ is defined by the condition $x=0$ and $x_{\perp}^i,\, i=1,..,d{-}1$ are the coordinates on $A_\perp$.\footnote{In this case, $A_\perp$ is a plane and the Cartesian coordinate $x$ is orthogonal to $A_\perp$.}

Here we consider the wavefunction for the vacuum state, which is built by performing the path integral over the lower half of the total Euclidean spacetime $(\tau\leq 0)$ such that the quantum field satisfies the
boundary condition $\psi(\tau=0,x,x_{\perp})=\psi_0(x,x_\perp)$ 
\begin{equation}
  \Psi[\psi_0(x,x_\perp)]=\int_{\psi(X)|_{\tau=0}=\psi_0(x,x_\perp)}
   {\rm D}\psi\; e^{-W[\psi]}~,
  \label{eq:wf_1}
 \end{equation}
where $W[\psi]$ is the action of the field.  The (co-dimension 2) surface $A_\perp$  separates the hypersurface $\tau=0$ into two parts:
$x<0, (A_c)$ and $x>0, (A)$. Thus the path integral boundary data $\psi(x,x_\perp)$ are split into 
\begin{align}
\psi(x,x_\perp)
=
\left\{
\begin{array}{llll}
\displaystyle
\psi_-(x,x_\perp)=\psi_0(x,x_\perp)&, \ \ \ & x<0 &,
\nonumber \\
\displaystyle
\psi_+(x,x_\perp)=\psi_0(x,x_\perp) &, \ \ \  & x>0& .
\end{array}
\right.  
\end{align}

The reduced density matrix describing the subregion $A$ ($x>0$) of the vacuum state is then obtained by tracing over the set of boundary fields $\psi_-$ located in the complementary region $A_c$. In the Euclidean path integral, this corresponds to integrating out $\psi_-$ over the entire spacetime, but with a cut from negative infinity to $A_\perp$ along the $\tau=0$ surface (i.e., along $x<0$). We must therefore impose boundary conditions for the remaining field $\psi_+$ as this cut is approached from above ($\psi_+^1$) and below ($\psi_+^2$). Hence we have:
 \begin{equation}
  \rho_A(\psi_+^1,\psi_+^2)=\int {\rm D}\psi_-\Psi(\psi_+^1,\psi_-)\Psi(\psi_+^2,\psi_-) \ .
 \label{eq:red_dm}
 \end{equation} 

Computing the von Neumann entropy $S_A=-\mathrm{Tr}\rho_A\log\rho_A$ from this formal object is an extremely difficult task for all but the very simple systems. The solution is given by the replica trick. The trace of the $n$-th power of the density matrix \eqref{eq:red_dm} is  given by the Euclidean path integral over fields defined on an $n$-sheeted covering  of the cut geometry associated to $\rho_A$. Taking polar coordinates $(r,\phi)$ in the $(\tau, x)$ plane, the cut corresponds to values $\phi=2\pi k,$
$k=1,2,..,n$. In building the $n$-sheeted cover, we glue sheets along the cut in such a way that the fields are smoothly continued from $\psi_+^{1,2}\big|_k$ to $\psi_+^{1,2}\big|_{k+1}$. The resulting space is a cone $C_n$, with angular deficit $2\pi(1-n)$ at $A_\perp$.  The partition function for the fields over this $n$-fold, which is denoted by  $Z[C_n]$, results
 \begin{equation}
  \Tr\rho_A^n=Z[C_n] \ .
 \label{eq:z_cone}
 \end{equation}
 
Assuming that in \eqref{eq:z_cone} we can consider an analytic continuation to
non-integer values of $n$, we have
\begin{align}
S_A=-{\rm Tr} \rho_A \log
\rho_A=-(\lambda\partial_\lambda-1)\log {\rm Tr}\rho_A^\lambda|_{\lambda=1}
\, .
\end{align}

Hence, introducing the effective action $W[\lambda]=-\log Z[C_\lambda]$ for fields on an Euclidean spacetime with a conical singularity at $A_\perp$, the cone $C_\lambda$ is defined, in polar coordinates, by making $\phi=\phi + 2\pi \lambda$. Then taking the limit in which $(1-\lambda)\ll1$, the entanglement entropy is given by the replica trick as 
\begin{equation}  
  S_A=(\lambda\partial_\lambda-1)W(\lambda)|_{\lambda=1}\, .
  \label{eq:ee_action}
\end{equation}
It is the action $W(\lambda)$ the function to be calculated. It can be shown that for a bosonic field whose partition function is $Z=\det^{-1/2}\cD$, with $\cD$ a differential operator, this action can be written as
\begin{align}
W=-\frac12\int_{\epsilon^2}^\infty \frac{ds}{s^2}\Tr{K(s)}
\ ,
\end{align}
with $K(s,X,X')$ the heat kernel satisfying
\begin{align}
(\partial_s +\cD)K(s,X,X')=0 
\ ,
\qquad\qquad
K(0,X,X')=\delta(X-X')
\ .
\end{align}
The heat kernel $K_\lambda(s,X,X')$ is obtained by applying the Sommerfeld formula \cite{sommerfeld}
\begin{align}
K_\lambda(s,X,X')
=
K(s,X,X')
+\Delta_\lambda(s,X,X')
\ ,
\end{align}
where $\Delta_\lambda$ ensures the $2\pi\lambda$ periodicity (see \cite{Solodukhin:2011gn} for further details).

\subsection*{Entanglement Entropy in Free Field Theory}
It can be proven that, for the operator $\cD=-\nabla^2+m^2$, one can obtain $K_\lambda(s,X,X')$ and then calculate $W(\lambda)$. The final result is
\begin{align}
\Tr{K_\lambda(s)}
=
\frac{1}{(4\pi s)^{d/2}}\left(
	\lambda V+2\pi \frac{1-\lambda^2}{6\lambda}s |A_\perp|
	\right)
\ ,
\end{align}
where $V$ is the spacetime volume and $|A_\perp|=\int d^{d-2}x$ is the area of the surface $A_\perp$.

On the other hand, the Euclidean path integral for the fields in a free theory is given by
\begin{align}
Z[J]=\int {\rm D}\psi\, e^{-W[\psi,J]}={\rm det}^{-1/2}(\cD)\,\exp\left(\frac{1}{4}J \cD^{-1} J\right)\, ,
\end{align}
where $J$ is the source for the matter field $\psi$, and $\cD$ is a differential operator related to the connected Green function by
\begin{align}
G(x,y)=\langle\psi(x)\psi(y)\rangle
=\frac{1}{Z[0]}\, \left. \frac{\delta^2 Z[J]}{\delta J(x)\delta J(y)}
\right|_{J=0}=\cD^{-1}
\, .
\end{align} 

After properly normalizing, we have 
\begin{align}
 W=\frac{1}{2}\log{\rm det}\, \cD = \frac{1}{2} {\rm Tr}\, \log (-\nabla^2+m^2)
 \, .
\end{align}

With this, $\Tr{K_{A_\perp}(s)}$, which is given by
\begin{align}
\Tr{K_{A_\perp}(s)}
=
\frac{|A_\perp|}{(4\pi s)^{\frac{d-2}{2}}}\exp(-m^2s)
\ ,
\end{align}
can be interpreted as the trace of the heat kernel of $\cD(A_\perp)$, where $\cD(A_\perp)$ is the differential operator over the codimension-2 plane $A_\perp$:
\begin{align}
\log\det\cD=-\int_{\epsilon^2}^\infty\frac{ds}{s}\Tr{K_{A_\perp}(s)}
\ .
\end{align}
Thus, upon a straightforward identification we obtain
\begin{align}
S_A = -\frac{1}{12}\, \log {\rm det}\,  \cD({A_\perp}) = -\frac{1}{12}\, {\rm Tr}_{A_\perp}\, \log\,  \cD
\, .
\label{eq:ee_half_space}
\end{align}

Based on this expression, in Sections \ref{sec:ee_cmera} and \ref{sec:holo} we will establish a relation between cMERA and the Ryu-Takayanagi formula.

\section{A cMERA Primer}\label{sec:cMERA}
cMERA \cite{Haegeman:2011uy, Nozaki:2012zj} amounts to a real space renormalization group procedure on the quantum state that builds, through a Hamiltonian evolution in scale,  scale dependent wavefunctionals $\Psi[\phi,u]$ given by,
\begin{equation}
\label{cMERAansatz}
\Psi[\phi,u]=\langle \phi|\Psi_u\rangle = \langle \phi|\, \mathcal{P} \, e^{-i \int_{u_{\text{IR}}}^u (K(u') + L) \, du' } \, |\Omega\rangle
\, .
\end{equation}
Here $u$ parametrizes the scale of the renormalization and $\mathcal{P}$ is the $u$-ordering operator.  $L$ represents the dilatation operator and $K(u)$ is the generator of evolution in scale, the so-called ``entangler'' operator. The scale parameter $u$ is taken to be in the interval $[u_{IR},u_{UV}] = (-\infty,0]$.  $u_{UV}$ is the scale at the UV cut off $\epsilon$, and the corresponding momentum space UV cut off is $\Lambda = 1/\epsilon$.  $u_{IR}$ is the scale in the IR limit.  

The state $|\Psi_{\Lambda}\rangle \equiv|\Psi_{ UV} \rangle$ is the state in the UV limit and it may be the ground state of a quantum field theory. The state $| \Omega \rangle$ is defined to have no entanglement between spatial regions.  $|\Omega \rangle$ is invariant with respect to spatial dilatations, so that  $e^{-iLu}|\Omega \rangle=|\Omega \rangle$ or, equivalently $L |\Omega \rangle=0$.

For a free bosonic theory, $|\Omega\rangle$ is defined by
\begin{align}
\label{scaleinv1}
\left(\sqrt{M} \,(\phi(\vk) - \bar \phi ) + \frac{i}{\sqrt{M}} \, \pi(\vk)\right) |\Omega\rangle = 0
\, ,
\end{align}
for all  momenta $\vk,$ where $M=\sqrt{\Lambda^2 + m^2}$ {with $m$ the mass of the particles in the free theory} and {$\bar \phi\equiv \matrixel{\Omega}{\phi(x)}{\Omega}$.} This state satisfies 
\begin{align}
\langle\Omega|\phi(\vp)\phi(\vq)|\Omega\rangle =\frac{1}{2 M}\ \delta^{d}(\vp+\vq)\, ,
\quad
\langle\Omega|\pi(\vp)\pi(\vq)|\Omega\rangle=\frac{M}{2}\ \delta^d(\vp+\vq)\, .
\end{align}

The nonrelativistic dilatation operator $L$ does not depend on the scale $u$ but only by the scaling dimensions of the fields. It is taken as the ``free'' piece of the cMERA Hamiltonian and is given by
\begin{align}
L=-\frac{1}{2}\int\, d\vx \left[\pi(\vx)\left(\vx \cdot \nabla\phi(\vx)\right) + \left(\vx \cdot \nabla\phi(\vx)\right)\pi(\vx) + \frac{d}{2}\left(\phi(\vx)\pi(\vx) + \pi(\vx)\phi(\vx)\right)\right]\, .
\end{align}

On the other hand, the entangler operator $K(u)$,  contains all the variational parameters to be optimized, creating entanglement between field modes with momenta $|\vk| < \Lambda$, where $\Lambda$ is the cutoff mentioned above. The entangler is considered as the ``interacting'' part of the cMERA Hamiltonian. From this point of view, the unitary operator in Eq. \eqref{cMERAansatz} 
\begin{equation}
U(u_1,u_2) \equiv \mathcal{P} \exp \left[ -i \int_{u_2}^{u_1} du \ (K(u) + L ) \right]\, 
\end{equation}
is understood as a Hamiltonian evolution with $K(u)+ L$ and thus it is useful to define cMERA in the ``interaction picture'' through the unitary transformation  $| \Phi_u \rangle  = e^{i L u} | \Psi_u\rangle$.

In this picture, the entangler is given by $\hat{K}(u) =  e^{i L \, u} \, K(u)  e^{-i L \, u}$, and the $u$-evolution is determined by the unitary operator
\begin{equation}\label{eq:cMERA_evol_free}
\cU(u_1,u_2) = \mathcal{P} \exp \left[ -i \int_{u_2}^{u_1} du \, \hat{K}(u) \right]\, .
\end{equation}

\subsection{Gaussian cMERA}
For free scalar theories in $(d +1)$ dimensions, $K(u)$ is given by the quadratic operator  \cite{Haegeman:2011uy,Nozaki:2012zj}
\begin{equation}
K(u) =
\frac{1}{2}\int_{\vp\vq}\, g(p;u)\, \left[
\phi(\vp)\pi(\vq)
+ \pi(\vp)\phi(\vq)
\right]\bar\delta(\vp+\vq)
\, ,
\label{eq:entangler}	
\end{equation}
{where $\int_{\mathbf{p}}\equiv\int\, (2\pi)^{-d}\, d^{d}\vp$.  The conjugate momentum of the field $\phi(\vp)$ is $\pi(\mathbf{p})$, such that $[\phi(\vp), \pi(\vq)] = i\bar{\delta}(\mathbf{\vp+\vq})\,$, with $\bar{\delta}(\vp)\equiv(2\pi)^{d}\delta(\vp)$. The function  $g(p;u)$ in (\ref{eq:entangler}) is the only variational parameter to be optimized in the cMERA circuit.} This function factorizes as 
\begin{equation}
\label{eq:free_var}
g(p;u)=g(u)\cdot\Gamma(p/\Lambda)\, ,
\end{equation} 
where  $\Gamma(x)$ is a cut off function which, in general, will be assumed $\Gamma(x)\equiv \Theta(1-|x|)$ with $\Theta(x)$ is the Heaviside step function. $g(u)$ is a real-valued function known as \emph{density of disentanglers} and $\Gamma(p/\Lambda)$ implements a high frequency cutoff such that $\int_{\vp} \equiv \int_0^{\Lambda}d^{d}\vp$ \cite{Haegeman:2011uy, Nozaki:2012zj}. The sharp cutoff function, which is assumed by default along this paper, ensures that $K(u)$ acts locally in a region of size $\epsilon \simeq \Lambda^{-1}$. 

In the interaction picture the entangler operator reads as follows:
\begin{align}
\hat{K}(u)
=\f{1}{2}\int_{\vp\vq} g(pe^{-u};u)\, \left[\phi(\vp)\pi(\vq)+\pi(\vp)\phi(\vq)\right]\bar\delta(\vp+\vq)\, 
\ .
\label{disein}
\end{align}

The optimized cMERA ansatz for the relativistic free massive scalar theory can be obtained as follows \cite{Haegeman:2011uy,Nozaki:2012zj}: The expectation value of the Hamiltonian of the theory w.r.t. the cMERA state $|\Psi_\Lambda\rangle$ is calculated in terms of the variational function $f(k,u_{\mathrm{IR}})$
\begin{align}
 f(k,u_{\mathrm{IR}})=\int^{u_{\mathrm{IR}}}_0 g(k e^{-u};u)\, du=\int^{-\log \Lambda/k}_0 g(u)\, du \ .
\end{align}
The optimization process implies
\begin{align}
 f(k,u_{\mathrm{IR}})=\frac{1}{4}\log \frac{k^2 + m^2}{M^2} \ .
\label{eq:f_free}
\end{align} 
This finally yields
\begin{align}
  g(u)=\frac{1}{2} \frac{\Lambda^2 e^{2u}}{(\Lambda^2 e^{2u}+m^2)}\, . \label{g_variational} 
 \end{align}

Remarkably, in \cite{Cotler:2016dha}, a cMERA circuit based on the quadratic entangler \eqref{eq:entangler} was used to study the self-interacting $\phi^4$ scalar theory. This model has a mass gap and flows to a free theory in the IR, where the IR ground state is exactly a Gaussian wavefunctional.  Similar to the free case, by minimizing the expectation value of the Hamiltonian with respect to the ansatz wavefunctional we obtain
\begin{align}
 f(k,u_{\mathrm{IR}})=\frac{1}{4}\log \frac{k^2 + \mu^2}{M^2}, \qquad
 k<\Lambda \ ,
 \label{eq:f_lambda}
 \end{align} 
and
\begin{align}
\label{gsol1}
g(k;u) =& \, \frac{1}{2} \, \frac{e^{2u}}{e^{2u} + \mu^2/\Lambda^2} \,  \Gamma(k/\Lambda)\, , 
\end{align}
where $\lambda$ is the coupling constant. In addition, $\mu$ is the modified mass of propagating free quasi-particles given by the \emph{gap} equation
\begin{equation}
\label{gapeqn1}
\mu^2 = m^2 +\frac{\lambda}{2}\left(\bar \phi^2+G(0)\right)\equiv m^2 + \frac{\lambda}{2}\, \Delta\, ,
\qquad
G(0)=\frac12 \int_\vk \frac{1}{\sqrt{k^2+\mu^2}}
\ .
\end{equation}

The cMERA wavefunctional thus obtained is a vacuum state for a free theory with mass given by \eqref{gapeqn1}.   The optimized ansatz captures all 1-loop $2$-point correlation functions, as well as the resummation of all cactus-like diagrams. 

\section{Entanglement Entropy in Gaussian cMERA}\label{sec:ee_cmera}
Here, we consider the entanglement entropy of half space in  free scalar theory in ($d+1$)-dimensions. In this case the entangling surface is $A_\perp = \mathbb{R}^{d-1}$ and its area will be denoted by $|A_\perp|$. According to the heat kernel result \eqref{eq:ee_half_space}, the entanglement entropy of the half space can be written as \cite{Hertzberg:2012mn,Fernandez-Melgarejo:2020utg}
\begin{align}
\label{eq:ee_inception}
S_A 
= \frac{|A_\perp|}{6}\, \int\, d^{d-1}\vk_\perp\, \log \expval{\Psi_\Lambda | \phi(\vk_\perp)\phi(-\vk_\perp)|\Psi_\Lambda}\, 
+\text{const}
\ ,
\end{align}
where $\text{const}$ represents a (UV dependent) quantity independent of mass and $|\Psi_\Lambda\rangle$ is the ground state of the field theory under consideration defined at some fixed cutoff $\Lambda$. The integration is carried out over the $(d-1)$ transverse momenta in $A_\perp$. Upon this assumption, from now on we will simplify the notation $\int\, d^{d-1} \vk_\perp \to \int d^{d-1} k $.

The Gaussian cMERA circuit introduced in the previous section is exactly solvable for the free scalar theory and thus, the UV cMERA approximation to the ground state of the theory $\ket{\Psi_\Lambda}$,  is  exact in this case. Upon these conditions, we compute the entanglement entropy of the half-space by renormalizing the 2-point correlator in \eqref{eq:ee_inception} with cMERA. First we note that,
\begin{dmath}
\expval{\Psi_\Lambda | \phi(\vk)\phi(-\vk)|\Psi_\Lambda}
=
\frac{1}{2M}\, e^{-2\,f(k,u_{\rm IR})}\, \delta^d(0)\, .
\end{dmath}
Therefore, using \eqref{eq:f_free} we express the entropy as
\begin{align}
S_A = \frac{|A_\perp|}{6} \int d^{d-1} k\, \log \left(\frac{e^{-2f(k,u_{\rm IR})}}{2M}\right) + {\widetilde{\rm const}}\, ,
\end{align}
where ${\widetilde{\rm const}}$ is a new UV dependent quantity independent of mass.

As noted above, when applying the Gaussian cMERA to the self interacting $\lambda\, \phi^4$ scalar theory \cite{Cotler:2016dha}, $f(k, u_{\rm IR})$ is given by
\begin{align}
f(k, u_{\rm IR})
=
\frac{1}{4}\log\frac{k^2+\mu^2}{M^2}
\ ,
\end{align}
where $\mu$ is the variational mass that satisfies the gap equation \eqref{gapeqn1}. 
Expanding for weak $\lambda$ one recovers the half space entropy for the theory at 1-loop \cite{Hertzberg:2012mn},
\begin{align}
S_A = -\frac{|A_\perp|}{12} \int d^{d-1} k\,  \left[\log \left(k^2+m^2\right)+\frac{\lambda}{2}\, \frac{\Delta}{\left(k^2+m^2\right)}\right]+ {\widetilde{\rm const}}
+\cO(\lambda^2)
\ .
\end{align}
This is precisely a consequence of the 1-loop exactitude of the Gaussian ansatz.

Let us now  write the half space entropy in cMERA in terms of the tensor network bulk variational parameters. To do this we note that
\begin{dmath}
\log \expval{\Psi_\Lambda | \phi(\vk)\phi(-\vk)|\Psi_\Lambda}
= 
- 2 f(k, u_{\rm IR})  + {\rm const}'
\, .
\end{dmath}
Consequently, we have
\begin{align}
\label{eq:ee_cmera_master}
S_A = -\frac{|A_\perp|}{3} \int d^{d-1}k\, f(k, u_{\rm IR})+\widetilde {\rm const}'\, .
\end{align}
Now, according to the definition of $f(k,u_{\rm IR})$, we have 
\begin{align}
 f(k, u_{\rm IR})= \int_0^{u_{\rm IR}}\, du\, g(k e^{-u},u)
 \, ,
 \qquad\qquad
g(k,u)\equiv g(u)\cdot \Gamma(k/\Lambda)\, .
\end{align}
Considering these quantities and taking into account that we are integrating over momenta $0\le k \le \Lambda$, we find
\begin{dmath}
S_A = \frac{|A_\perp|}{3} \int d^{d-1}k  \int_{u_{\rm IR}}^0\, du\, g(k e^{-u},u) +\widetilde {\rm const}'
=\frac{|A_\perp|}{3}  \int^0_{u_{\rm IR}}\, du\, g(u)\,  \int\, d^{d-1}k\, k^{d-2} \, \Gamma(k e^{-u}/\Lambda) +\widetilde {\rm const}'
\ .
\end{dmath}
We note that this expression is also valid for other choices of the cutoff function $\Gamma(k/\Lambda)$. Namely one can rewrite the last expression as
\begin{align}
S_A=\frac{|A_\perp|}{3}  \int^0_{u_{\rm IR}}\, du\, g(u)\cdot  \Sigma(u) +\widetilde {\rm const}'
\ ,
\qquad
\Sigma(u)\equiv \int\, d^{d-1}k \, k^{d-2} \, \Gamma(k e^{-u}/\Lambda)
\ .
\label{eq:Sigma}
\end{align}

\begin{table}
\begin{align*}
\begin{array}{|c|c|c|}
\hline
\Gamma\left(k/\Lambda\right) & \Sigma(u) & \frac{4G_N}{\gamma}\sqrt{G_{uu}}/g(u)
\\
\hline
\displaystyle \Theta\left(1-\frac{k e^{-u}}{\Lambda}\right) & \displaystyle \frac{S^{(d-1)}}{d-1} \Lambda^{d-1}e^{u(d-1)} & 1
\\[1em]
\displaystyle \frac{1}{2} \exp \left(-\frac1\sigma \frac{k^2 e^{-2 u}}{\Lambda ^2}\right) &  \frac{S^{(d-1)}}{4}  \sigma ^{\frac{d-1}{2}}\, \Gamma \left(\frac{d-1}{2}\right)\, \Lambda ^{d-1}\, e^{u(d-1)}  &  \frac{1}{4} \, \sigma ^{\frac{d-1}{2}}\,  \Gamma \left(\frac{d-1}{2}\right)
\\[1em]
\displaystyle \frac12 \frac{\Lambda^2}{\Lambda^2+k^2 e^{-2 u}} &  \frac{S^{(d-1)}}{4}\,  \pi \,  \Lambda ^{d-1} e^{u(d-1)}\, \, (d=2) & \frac{\pi}{4} 
\\
\hline
\end{array}
\end{align*}
\caption{\textit{For various cutoff functions $\Gamma(k/\Lambda)$, their corresponding momentum integral $\Sigma(u)$ defined in \protect\eqref{eq:Sigma} and the ratio between the radial AdS metric component and the variational parameter, $\sqrt{G_{uu}}/g(u)$ are shown. $S^{(d)}$ is the area of a unit $d$-sphere, $S^{(d)} \equiv \frac{2\pi^{d/2}}{\Gamma(d/2)}$.}}
\label{tab:cutoffs}
\end{table}

The cut off functions, which act as alternative UV regularization schemes of the cMERA formalism, will determine $\Sigma(u)$. In Table \ref{tab:cutoffs}, it is shown the resulting $\Sigma(u)$ for some cutoff functions that have been proposed in the literature \cite{Haegeman:2011uy,Zou:2019xbi}.  We observe that the result is similar to the one obtained through sharp cut off function up to some numerical factors. Consequently, and for convenience, in the rest of this section we will consider the sharp cutoff  $\Gamma(x)=\Theta(1-x)$. In this case the entanglement entropy results
\begin{dmath}
S_A 
=
\frac13\frac{S^{(d-1)}}{(d-1)} |A_\perp|   \Lambda^{d-1}\, \int^0_{u_{\rm IR}}\, du\, g(u)\,  e^{u(d-1)} +\widetilde {\rm const}'
\ ,
\end{dmath}
with $S^{(d)}$ the area of a $d$-sphere, $S^{(d)} \equiv \frac{2\pi^{d/2}}{\Gamma(d/2)}$.

Thus the final result can be written as
\begin{align}
\label{eq:ee_cmera_boxed}
S_A=\frac13\frac{S^{(d-1)}}{(d-1)} \frac{|A_\perp|}{\epsilon^{d-1}}\, \int^0_{u_{\rm IR}}\, du\, g(u)\,  e^{u(d-1)} +\widetilde {\rm const}'\, .
\end{align}
where $\epsilon \equiv 1/\Lambda$ can be identified as the lattice constant.

Let us compare this expression with the half space entanglement entropy $S_A$ in the discrete MERA for a $(d+1)$-dimensional quantum system on a lattice which was obtained in \cite{Nozaki:2012zj}. In this case the entanglement entropy is given by
\begin{align}
  S_A\propto L^{d-1}\sum_{u=-\infty}^0 n(u)\cdot 2^{(d-1)u}\, ,
  \label{eq:ee_dmera}
\end{align}
where $L^{d-1}$ is the number of lattice points on the boundary of $A$ and $n(u)$
is the strength of the bonds at the layer specified by the non-positive integer $u$. $n(u)$ is typically given by the logarithm of the bond dimension at layer $u$. The comparison between these expressions shows a manifestly similar structure and suggests that $g(u)$ in cMERA might be interpreted as a local \emph{bond dimension}. We will elaborate on this in Secs. \ref{sec:holo} and \ref{sec:discussion}.

So far, we have obtained the same quantity, $S_A$, as an integral over momenta, Eq. \eqref{eq:ee_inception} and over the scale, Eq. \eqref{eq:ee_cmera_boxed}. Upon considering the integrand of the former expression, we define the entanglement entropy at the scale $u$, $S_A(u)$, as
\begin{align}
\begin{split}
S_A(u)
\equiv&\ \frac{|A_\perp|}{6}\, \int_0^{\Lambda e^u}\, d^{d-1}k_\perp\, \log \matrixel{\Psi_\Lambda }{ \phi(\vk_\perp)\phi(-\vk_\perp)}{\Psi_\Lambda}\, 
+\text{const}
\\
=&\ \frac{|A_\perp|}{6}\, \int_0^{\Lambda}\, d^{d-1}k_\perp\, e^{u(d-1)}\log \matrixel{\Psi_\Lambda }{ \phi(\vk_\perp e^{u})\phi(-\vk_\perp e^{u})}{\Psi_\Lambda}\, 
+\text{const}
\ .
\end{split}
\label{eq:ee_u_dep}
\end{align}

\begin{figure}[!t]
\begin{center}
\includegraphics[height=.34\textheight]{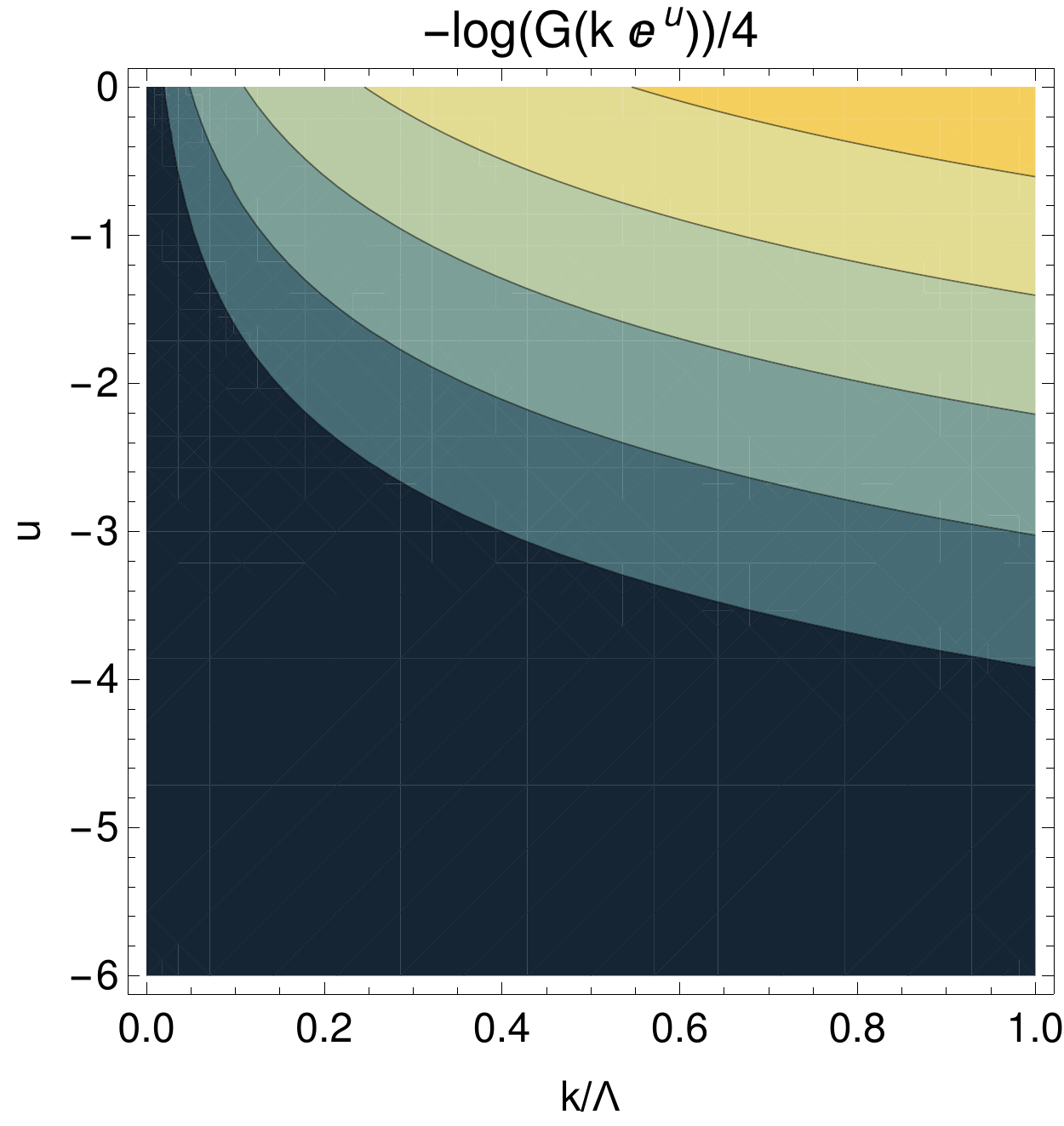}
\includegraphics[height=.345\textheight]{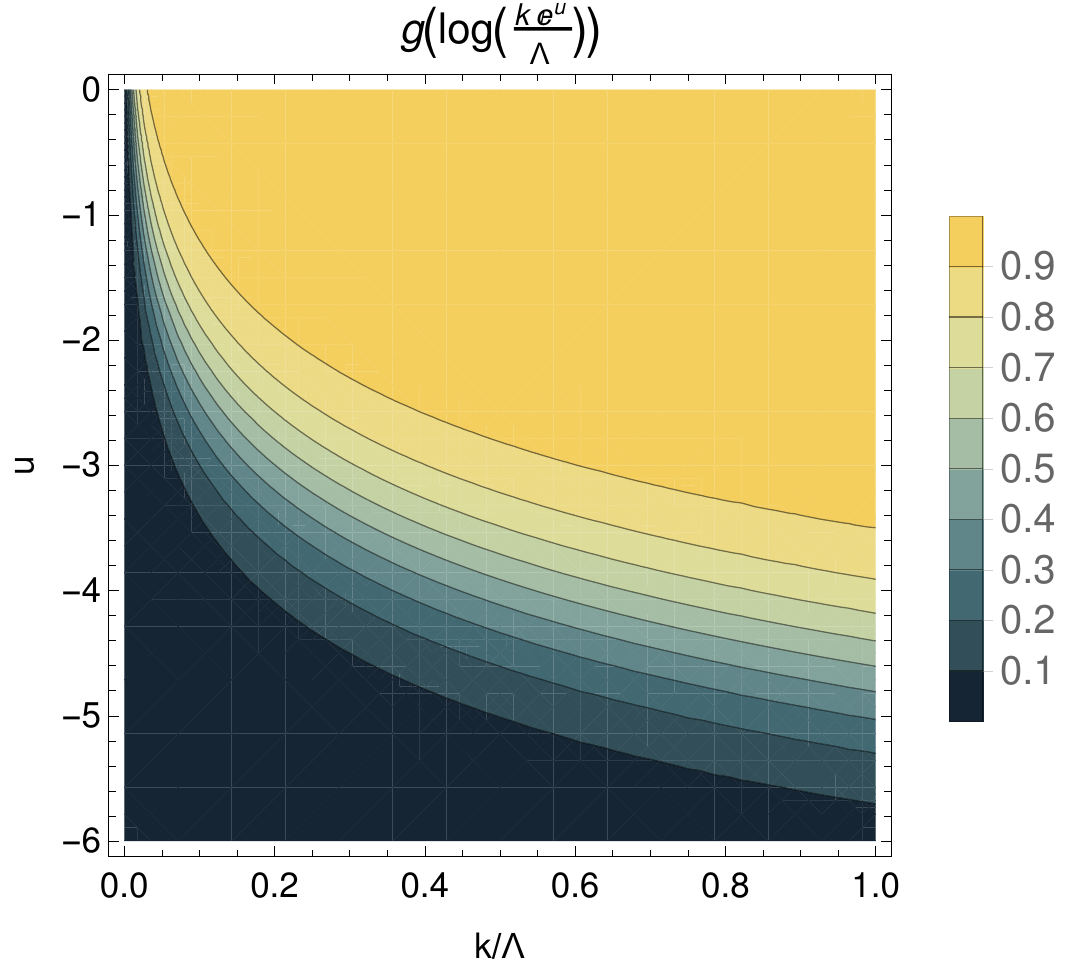}
\caption{\textit{Left: Rate of entanglement entropy per unit momentum (\emph{c.f.}  Eq. \protect\eqref{eq:ee_cmera_boxed}).  Right: Rate of entanglement entropy per unit scale (\emph{c.f.}  Eq. \protect\eqref{eq:ee_u_dep}).  Parameters: $d=1$, $\Lambda=100$, $m=\frac{\Lambda}{100}$. }}
\label{fig:integrands}
\end{center}
\end{figure}

To gain some insight into the entropy production rate as a function of the momentum and as a function of the scale, let us study  the integrands of \eqref{eq:ee_cmera_boxed} and \eqref{eq:ee_u_dep} 
\footnote{Let us note that, because we are introducing some momentum cutoffs, $\text{const}$ in \eqref{eq:ee_u_dep} is just a finite UV cut off dependent quantity.}, as both procedures must give the same result. In Fig. \ref{fig:integrands} we plot, for $d=1$, the log of the 2-point correlator $G(k_\perp e^u)\equiv \matrixel{\Psi_\Lambda }{ \phi(k_\perp e^{u})\phi(-k_\perp e^{u})}{\Psi_\Lambda}$ from \eqref{eq:ee_u_dep} and the variational parameter $g(\log(k e^u/\Lambda))$ from \eqref{eq:ee_cmera_boxed}. For the latter, a change of variable is needed to consistently compare the two definite integrals. We observe that, qualitatively, both of them exhibit an area law behavior: highest momenta, \emph{i.e.} short distance correlations, provide the highest contributions to the entropy. In contrast, while in the left plot higher momenta are gradually incorporated as $u\to0$, in the right plot all momenta are contributing at relatively low $u$ and only the renormalization scale determines the strength of the entanglement.

\section{cMERA and Holography}
\label{sec:holo}

In order to establish an explicit connection between cMERA and holography, it would be desirable to define local areas in the bulk in terms of the differential entanglement generated along the tensor network, as it was suggested in \cite{Swingle:2012wq}. That is to say, the infinitesimal contribution to the area of an hypothesized minimal surface homologous to the half space, must be defined to be proportional to the change in entropy. From \eqref{eq:ee_cmera_boxed}, the differential entropy production rate is given by 
\begin{align}
\frac{d S_A}{du}=\gamma\, \frac{|A_\perp|}{\epsilon^{d-1}}\,  g(u)\,  e^{u(d-1)}
\, ,
\label{eq:ee_cmera_diff}
\end{align} 
where $\gamma \equiv \frac13\frac{S^{(d-1)}}{(d-1)}$.  With this, the proposal amounts to recovering a dual tensor network bulk metric as an inverse problem in terms of this differential entropy. It is also convenient to note that in the free boson theory, where $0 \leq g(u) \leq 1$, the differential form of our result \eqref{eq:ee_cmera_diff} fulfills the following inequality
 \begin{align}
\frac{d S_A}{du} \leq \gamma\, \frac{|A_\perp|}{\epsilon^{d-1}}\, e^{u(d-1)} \, .
\end{align} 
This upper bound on the entanglement entropy along a cMERA flow, which was given in \cite{Haegeman:2011uy}, is based solely on very general conditions for the Hamiltonian evolution in scale implemented by cMERA.

Given the results shown above,  we are in a position to establish a relation between the AdS geometry and the cMERA tensor network through the Ryu-Takayanagi formula. To this end, let us consider a Gaussian cMERA for $N$ free scalar fields with $O(N)$ symmetry. In this case the cMERA entangler is given by $N$ copies of the entangler for a free boson theory. With this, it is straightforward to conclude that the half space entropy amounts to
\begin{align}
S_A=  N\, \gamma\, \frac{|A_\perp|}{\epsilon^{d-1}}\, \int_{u_{\rm IR}}^0\, du\, g(u)\,  e^{u(d-1)} +\widetilde {\rm const}'\, .
\label{eq:cmera_ee_N}
\end{align}
For convenience, we write \eqref{eq:cmera_ee_N} as
\begin{align}
d S_A = N\, \gamma\cdot dA_{\text{TN}}\, ,
\label{eq:dee_cmera_c}
\end{align}
where
\begin{align}
dA_{\text{TN}} \equiv \frac{|A_\perp|}{\epsilon^{d-1}}\, g(u)\,  e^{u(d-1)}\, du
\, .
\label{eq:darea_cmera}
\end{align}
We understand $dA_{\text{TN}}$ as an infinitesimal area surface in the bulk of the tensor network. To see this, let us assume that an ansatz for a geometric description of the tensor network is given as the spatial slice of an asymptotically $(d +2)$-dimensional AdS metric  with a radial coordinate labeled by the cMERA parameter $u$:
 \begin{align}
 ds^2 = G_{uu}\, du^2 + \frac{e^{2u}}{\epsilon^2}\, d\vx_{d}^2 + G_{tt}dt^2\, .
 \end{align}
Here, the AdS radius has been fixed to be unity for simplicity and $G_{uu} \to {\rm constant}$ as $u \to 0$. On the other hand, the Ryu-Takayanagi formula for the half space in this geometry implies that \cite{Nozaki:2012zj} 
\begin{align}
 \label{eq:dee_rt}
 d\, S_A=\frac{1}{4 G^{(d+2)}_N}\cdot  d{\rm A}_{\text{RT}} \, ,
\end{align}
where
\begin{align}
d {\rm A}_{\text{RT}} \equiv \frac{|A_\perp|}{\epsilon^{d-1}}\, \sqrt{G_{uu}}\,  e^{u(d-1)}\, du
\label{eq:darea_rt}
\end{align}
is an infinitesimal area in the bulk geometry.

When comparing \eqref{eq:darea_cmera} and \eqref{eq:darea_rt} we obtain that the bulk geometry may be described in terms of the variational parameters of the tensor network. This definition arises from a computation of the entanglement entropy in cMERA following a QFT prescription and comparing the result with the holographic calculation. As a result, the entropic RG flow generated by cMERA can be consistently encoded in terms of an AdS geometry as far as the radial component of the metric is related to the variational parameter of the tensor network as
\begin{align}
\frac{1}{4G_N} \sqrt{G_{uu}} = \gamma\, g(u)
\, .
\end{align}
Upon identifying $\gamma=\frac{1}{4G_N}$, we give the ratio $\sqrt{G_{uu}}/g(u)$ for other alternative cutoff functions $\Gamma(k/\Lambda)$ in the last column of Table \ref{tab:cutoffs}.  It is straightforward to note that  different cMERA UV regularization schemes implemented by cutoff functions $\Gamma(ke^{-u}/\Lambda)$ do not affect the dual bulk metric up to order one numerical factors. In addition, as noted in \cite{Nozaki:2012zj}, for any cutoff function, changing it  by $\Gamma(k\eta(e^{-u})/\Lambda)$ just amounts to a coordinate transformation along the AdS radial direction of the form $e^{-u} \to \eta(e^{-u})$.

\subsection{Fisher Information Metric}
In \cite{Nozaki:2012zj}, in an attempt to make the connection between cMERA and the AdS/CFT more precise, authors showed that it is reasonable to think about the density of (dis)entanglers in terms of quantum distances. That is to say, characterizing how the entanglers entering the cMERA circuit modify a state at an infinitesimal scale $du$ would help to elucidate what is the entropy production rate. For this purpose, authors considered  two pure quantum states described by $\ket{\psi_1}$ and $\ket{\psi_2}$. The Hilbert-Schmidt distance given by
\begin{align}
D_{\rm HS}(\psi_1,\psi_2)
\equiv
  1 - | \braket{\psi_1}{\psi_2}|^2\, ,
\end{align}
determines how different these states are. For a set of pure states $\left\{\ket{\psi(\lambda)}\right\}$ which are described by a set of parameters $\lambda=\{\lambda_1,\lambda_2,\ldots\}$, the Hilbert-Schmidt distance between two infinitesimally close states is given by
\begin{align}
D_{\rm HS}\left(
	\psi(\lambda),\psi(\lambda+\delta\lambda)
	\right)
= g_{ij}(\lambda)\delta\lambda_i\delta\lambda_j
\ , 
\end{align}
where $g_{ij}(\lambda)$ is the so called Fisher information metric and is given by
\begin{align}
g_{ij}(\lambda)
=
{\rm Re}\braket{\partial_i\psi(\lambda)}{\partial_j\psi(\lambda)}
-\braket{\partial_i\psi(\lambda)}{\psi(\lambda)}\braket{\psi(\lambda)}{\partial_j\psi(\lambda)}
\ .
\end{align}
Following these definitions, a Fisher metric for the cMERA $g_{uu}$ can be defined as
\begin{align}
g_{uu}(u)du^2
=
\mathcal{N}^{-1}\left(
	1-
	\left|\braket{\Phi(u)}{\Phi(u+du)}\right|^2
	\right)
\ ,
\qquad\quad
\ket{\Phi(u)}\equiv P e^{-i\int_{u_{IR}}^u \hat K(s)ds}\ket{\Omega}
\ ,
\end{align}
where $\mathcal{N}$ is a normalization constant. The result is 
\begin{align}
g_{uu}(u)
=
\matrixel{\Phi(u)}{\hat K(u)^2}{\Phi(u)}
-\matrixel{\Phi(u)}{\hat K(u)}{\Phi(u)}^2
\ .
\end{align}
For the particular case of the Gaussian entangler \eqref{eq:entangler}, the metric results $g_{uu}=g(u)^2$.

Therefore, because $g_{uu}$ measures the density of disentanglers \cite{Nozaki:2012zj}, the entanglement entropy obtained from the formula \eqref{eq:ee_cmera_boxed} can be naturally interpreted as the summation of the entanglers that cut the curve that divides the system at a certain scale $u$. 

In view of these results, here we note that in case of enlarging the number of entangler operators in a cMERA circuit,
\begin{align}
\hat K(u)=\hat K_1(g_1(u);u)+\ldots+\hat K_n(g_n(u);u)
\ ,
\label{eq:sum-entanglers}
\end{align}
with $\hat K_i$ an entangler containing a generic dependence on the scale parameters (dilatations) and a variational parameter $g_i(u)$,
the Fisher metric $g_{uu}$ will be a quadratic function of the variational parameters:
\begin{align}
g_{uu} = \sum_{ij} \alpha_{ij} g_i(u) g_j(u)
\ ,
\label{eq:sum-metric}
\end{align}
where $\alpha_{ij}$ are real valued coefficients associated to the expectation values of the multiple (products of) disentanglers. This suggests that adding more entanglers, and consequently having more terms in this sum, is the analog of increasing the bond dimension in the discrete MERA circuit. In this respect, some recent non-Gaussian formulations of cMERA for interacting field theories, incorporate additional non quadratic entanglers (with their respective variational parameters), giving rise to genuinely non-Gaussian effects \cite{Cotler:2016dha,Fernandez-Melgarejo:2019sjo,Fernandez-Melgarejo:2020fzw}. 

On the other hand, recent results in the literature \cite{Chen:2020ild,Fernandez-Melgarejo:2020utg,Iso:2021vrk} show how to obtain the entanglement entropy for non-Gaussian states in some particular cases. The gist of these formulations consists of replacing the Gaussian 2-point correlators appearing in the computation of the entanglement entropy in the free case by the same 2-point correlators evaluated on the non-Gaussian states.

Consequently, we conjecture that, in those cases that match the conditions imposed for the formulations commented above, the entanglement entropy for a cMERA circuit based on an entangler of the form \eqref{eq:sum-entanglers} is given by
\begin{align}
\label{eq:ee_cmera_guu}
S_A=\frac13\frac{S^{(d-1)}}{(d-1)} \frac{|A_\perp|}{\epsilon^{d-1}}\, \int^0_{u_{\rm IR}}\, du\, \sqrt{g_{uu}(u)}\,  e^{u(d-1)} +\widetilde {\rm const}'\, .
\end{align}
where the factor $\sqrt{g_{uu}}$, with $g_{uu}$ given by \eqref{eq:sum-metric}, is
the integral measure of the ``curved'' tensor network bulk space along the entanglement renormalization direction. Let us mention that this result is trivially in agreement with the case of $N$ Gaussian free fields. Some particular non-trivial cases are treated in detail in \cite{tbp}.

\section{Discussion}\label{sec:discussion}

The computation of the half space entanglement entropy for free scalar fields  through the replica trick and the heat kernel method shows that the entanglement entropy reduces to the momentum integral of a single 2-point correlator and satisfies the area law \eqref{eq:ee_inception}. 

Upon taking this expectation value as evaluated by the cMERA UV state,
we have obtained Eq. \eqref{eq:ee_cmera_boxed}. This expression entitles the variational parameter $g(u)$ to be (up to some factors) the differential entanglement generator at a certain scale $u$. 

To establish a one-to-one correspondence between the integrands of both expressions, we have introduced the entanglement entropy at scale $u$, $S_A(u)$, by considering a scale dependent momentum cut off $\Lambda e^{u}$, see \eqref{eq:ee_u_dep}. The entanglement production rate as a function of the momentum and the renormalization scale is shown in Fig. \ref{fig:integrands}, where we observe a similar qualitative behavior between the two integrands, together with an explicit identification of the area law due to the contribution of highest momenta. From these results, we find that the differential entropy production \eqref{eq:ee_cmera_diff} is in agreement with the upper bounds existing in the literature \cite{Haegeman:2011uy}. 

In Sec. \ref{sec:holo} we have studied our results from a holographic viewpoint.
We have shown a one-to-one correspondence between the infinitesimal area surface in the bulk of the tensor network $dA_{\text{TN}}$ and the Ryu-Takayanagi area surface $dA_{\text{RT}}$. We have explicitly found that, up to overall factors that could be relevant for the $G_N$ identification, the radial component of the AdS metric is proportional to the cMERA variational parameter $\sqrt{G_{uu}}\propto g(u)\cdot\Sigma(u)$, where $\Sigma(u)$ is the integral over momenta of the cutoff function $\Gamma(k e^{-u}/\Lambda)$. This result motivates the following two analyses.

Firstly, we have checked that, up to overall factors, different cMERA regularization schemes do not modify the dual metric in the bulk. This means that our interpretation of the cMERA differential entropy in terms of dual minimal area surfaces is completely decoupled from arbitrary choices of the UV regularization scheme of the tensor network. Secondly, we have noticed that having a factor $g(u)$ in the entanglement entropy just reduces to treating the simplest case in \eqref{eq:sum-metric}, namely a single or $N$ non interacting free scalar fields. 

Finally, given some new results on the entanglement entropy of interacting field theories \cite{Fernandez-Melgarejo:2020utg}, \cite{Chen:2020ild} as well as a recent proposal of cMERA circuits for interacting fields \cite{Fernandez-Melgarejo:2020fzw}, we conjecture that the entanglement entropy in theories with arbitrary values of the interaction coupling, could be non-perturbatively computed in cMERA through the Fisher information metric $g_{uu}$, the avatar of the bond dimension in the discrete version of the MERA circuit, by means of \eqref{eq:ee_cmera_guu}, where $\sqrt{g_{uu}}$ acts precisely as the integral measure of the ``curved'' tensor network bulk space along the scale direction. This will be explored in \cite{tbp}.

\section*{Acknowledgments}
We thank Esperanza L\'opez for very fruitful discussions and a careful reading that helped to improve the manuscript. The work of JJFM is supported by Universidad de Murcia-Plan Propio Postdoctoral, the Spanish Ministerio de Econom\'ia y Competitividad and CARM Fundaci\'on S\'eneca under grants FIS2015-28521 and 21257/PI/19. JMV is funded by Ministerio de Ciencia, Innovaci\'on y Universidades PGC2018-097328-B-100 and Programa de Excelencia de la Fundaci\'on S\'eneca Regi\'on de Murcia 19882/GERM/15.

\small


\providecommand{\href}[2]{#2}\begingroup\raggedright\endgroup

\end{document}